\documentclass[floatfix, noshowpacs, preprintnumbers, twocolumn, amsmath, amssymb, aps, prx, superscriptaddress]{revtex4-1}

\usepackage{graphicx, xcolor}
\usepackage{soul}
\usepackage{times}
\usepackage{multirow}
\usepackage{booktabs}
\usepackage{hhline}
\usepackage{rotating}
\usepackage{complexity}
\usepackage{color}
\usepackage{hyperref}
\usepackage{booktabs}
\usepackage{colortbl}
\usepackage{mathtools}
\usepackage[fleqn]{nccmath}
\usepackage{enumitem}
\usepackage{dsfont}
\usepackage{placeins}

\definecolor{Gray}{gray}{0.9}

\begin{document}

\title{Validating multi-photon quantum interference with finite data}

\author{Fulvio Flamini}
\affiliation{Dipartimento di Fisica, Sapienza Universit\`{a} di Roma, Piazzale Aldo Moro 5, I-00185 Roma, Italy}
\affiliation{Institut f\"{u}r Theoretische Physik, Universit\"{a}t Innsbruck, Technikerstra{\ss}e 21a, 6020 Innsbruck, Austria}

\author{Mattia Walschaers}
\affiliation{Laboratoire Kastler Brossel, Sorbonne Universit\'{e}, CNRS, ENS-PSL Research University, Coll\`{e}ge de France, 4 place Jussieu, 75005 Paris, France}

\author{Nicol\`o Spagnolo}
\affiliation{Dipartimento di Fisica, Sapienza Universit\`{a} di Roma, Piazzale Aldo Moro 5, I-00185 Roma, Italy}

\author{Nathan Wiebe}
\affiliation{Station Q Quantum Architectures and Computation Group, Microsoft Research, Redmond, WA, United States}
\affiliation{Pacific Northwest National Laboratory, Richland, WA, United States}
\affiliation{Department of Physics, University of Washington, Seattle, WA, United States}

\author{Andreas Buchleitner}
\affiliation{Physikalisches  Institut,  Albert-Ludwigs-Universit\"{a}t  Freiburg, Hermann-Herder-Strasse  3,  79104  Freiburg,  Germany}

\author{Fabio Sciarrino}
\affiliation{Dipartimento di Fisica, Sapienza Universit\`{a} di Roma, Piazzale Aldo Moro 5, I-00185 Roma, Italy}


\begin{abstract}
Multi-particle interference is a key resource for quantum information processing, as exemplified by Boson Sampling. Hence, given its fragile nature, an essential desideratum is a solid and reliable framework for its validation. However, while several protocols have been introduced to this end, the approach is still fragmented and fails to build a big picture for future developments.
In this work, we propose an operational approach to validation that encompasses and strengthens the state of the art for these protocols. To this end, we consider the Bayesian hypothesis testing and the statistical benchmark as most favorable protocols for small- and large-scale applications, respectively.
We numerically investigate their operation with finite sample size, extending previous tests to larger dimensions, and against two adversarial algorithms for classical simulation: the Mean-Field sampler and the Metropolized Independent Sampler.
To evidence the actual need for refined validation techniques, we show how the assessment of numerically simulated data depends on the available sample size, as well as on the internal hyper-parameters and other practically relevant constraints. Our analyses provide general insights into the challenge of validation, and can inspire the design of algorithms with a measurable quantum advantage.
\end{abstract}

\maketitle


\section{Introduction}

A quantum computational advantage occurs when a quantum device starts outperforming its best classical counterpart on a given specialized task \cite{Harrow17, Arute19}. Intermediate models \cite{AA, Terhal04, Bremner10, Boixo18} and platforms \cite{Inagaki16, McMahon16, Goto16, Lechner15, Johnson11, Puri17} have been proposed to achieve this regime, largely reducing the physical resources required by universal computation. The technological race towards quantum computational advantage goes nonetheless hand-in-hand with the development of classical protocols capable to discern genuine quantum information processing \cite{Shin14, Fitzsimons17, Flamini18, Eisert19, Walschaers19}. The intertwined evolution of these two aspects has been highlighted in particular by Boson Sampling \cite{AA, Brod19}, where several protocols have been introduced \cite{Mayer11, Aaronson14, Spagnolo14, Carolan14, Bentivegna14, Tichy10, Tichy14, Crespi15, Walschaers16, Liu16, Shchesnovich16, Wang16bubbles, Dittel17, Dittel18, Viggianiello18, Agresti19, Flamini19tsne} and experimentally tested \cite{Bentivegna14, Spagnolo14, Carolan14, Carolan15, Crespi16, Wang16beating, Wang16ten-photon, He17, Wang17, Viggianiello17tvd, Viggianiello18, Wang18, Giordani18, Agresti19, Paesani19, Wang1920photons} to rule out non-quantum processes.  Boson Sampling, in its original formulation \cite{AA}, consists in sampling from a probability distribution that can be related to the evolution of indistinguishable photons in a linear-optical interferometer. Recent analyses suggested reasonable thresholds in the number of photons $n$ to surpass classical algorithms \cite{Neville17, Clifford18, Wu16}. 

While the sampling task itself has been thoroughly analyzed in computational complexity theory, we still lack a comparable understanding when it comes to its validation. However, it is clear from a practical perspective that any computational problem designed to demonstrate quantum advantage needs to be formulated together with a set of validation protocols which account for the physical ramifications and resources required for its implementation. For instance, while small-scale examples can be validated by direct solution of the Schr\"{o}dinger equation and using statistical measures such as cross-entropy \cite{Boixo18}, this is prohibitively expensive to debug a faulty Boson Sampler. Moreover, for Boson Sampling a deterministic certification is impossible \cite{Tichy10} by the very definition of the problem \cite{Aaronson14}.
Hence, it is crucial to develop debugging tools, as well as tests to exclude undesired hypotheses on the system producing the output, that are computationally affordable and experimentally feasible. Furthermore, due to random fluctuations inherent to any finite-size problem, a validation cannot be considered reliable until sufficient physical resources are spent to obtain reasonable experimental uncertainties. Ultimately, no computational problem can provide evidence of quantum advantage unless quantitative validation criteria can be stated. 
 
In this work, we investigate the problem of validating multi-photon quantum interference in realistic scenarios with finite data.
The paper is structured as follows: first, we discuss possible ambiguities in the validation of Boson Sampling, which play a crucial role in large-size experiments. Then, building upon state-of-the-art validation protocols, we address the above considerations with a more quantitative analysis. We describe a practical approach to validation that makes the most of the limited physical resources available. Specifically, we study the use of the statistical benchmark \cite{Walschaers16} and the Bayesian hypothesis testing \cite{Bentivegna14} to validate $n$-photon interference for large and small $n$, respectively. We numerically investigate their operation against classical algorithms to simulate quantum interference, with a particular focus on the number of measurements.
The reported analysis strengthens the need for a well-defined approach to validation, both to demonstrate quantum advantage and to assist applications that involve multi-photon states.

\section{Validation of Boson Sampling: framework}

\label{sec:framework}

\begin{figure}[t]
\includegraphics[width=0.95\linewidth]{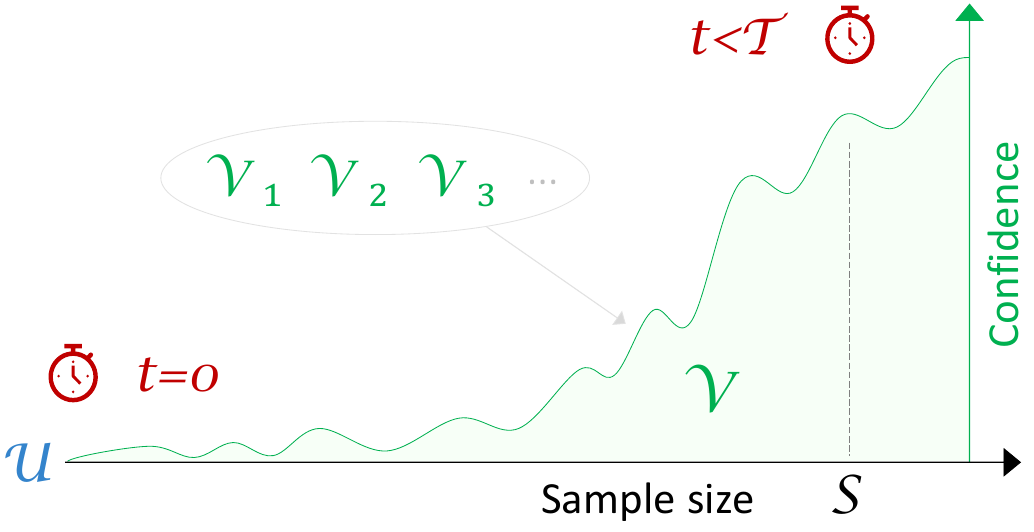}
\caption{To demonstrate quantum advantage, reliable and realistic approaches to validation need to be defined. Boson Sampling should be validated with well-defined sampling time ($\mathcal{T}$) and sample size ($\mathcal{S}$), since the efficacy of validation protocols ($\mathcal{V}$) changes with the number of measured events after the unitary evolution ($\mathcal{U}$).
}
\label{Fig1}
\end{figure}

Our aim, in the context of Boson Sampling, consists in the unambiguous identification of a quantum advantage in a realistic scenario. We focus on the task of \emph{validation}, or verification, whose aim is to check if measured experimental data is compatible with what can be expected from a given physical model. Validation generally requires fewer resources and is, thus, more appropriate for practical applications than full certification, which is exponentially hard in $n$ for Boson Sampling \cite{Aaronson14, Hangleiter19}. In both cases, these claims must follow a well-defined protocol to distill experimental evidence that is accepted by the community under jointly agreed criteria \cite{Markov18} (Fig. \ref{Fig1}). As we discuss below and in Sec. \ref{sec_partII}, we propose an application-oriented approach to validation that takes into consideration the limited physical resources, be them related to the evaluation of permanents \cite{Valiant79} or to finite sample size \cite{Hangleiter19}. In fact, without such well-defined approaches, obstacles or ambiguities may arise in large-scale experiments, as we highlight in the following. For instance, not all validation protocols are computationally efficient, which is a strong limitation for future multi-photon applications or high-rate real-time monitoring. Also, a theoretically scalable validation protocol may still be experimentally impractical due to large instrumental overheads or large prefactors that enter the scaling law. 

Given two validation protocols $\mathcal{V}_1$ and $\mathcal{V}_2$ to rule out the same physical hypothesis or model, which conclusion can be drawn if they agree for a data set of given size and unexpectedly disagree when we add more data?
In principle we can accept or reject a data set when we reach a certain level of confidence, but which action is to be taken if this threshold is not reached after a large number of measurement events (which hereafter we refer to as the ``sample size")? Shall we proceed until we pass that level, shall we reject it or shall we make a guess on the available data? Finally, what if the classical algorithm becomes more effective in simulating Boson Sampling for larger data sets, as for Markov chains \cite{Neville17}, or for longer processing times, as for adversarial machine learning algorithms \cite{adversarial-ML} that could exploit specific vulnerabilities of validation protocols?

However artificial some of the above questions may seem, such skeptical approach was indeed already adopted \cite{Tichy14} and addressed \cite{Dittel17, Dittel18, Crespi15, Viggianiello18, Flamini19tsne, Crespi16, Carolan15, Walschaers16} with the Mean-Field sampler (see Sec. \ref{Appendix_MF_Markov}): all these considerations are necessary to strengthen the claim of quantum advantage.
Under the above premise, we therefore identify the following crucial features to be assessed in any decision on acceptance or rejection:

\begin{enumerate}[leftmargin=*, align=left]
\item \textit{Sample size $\mathcal{S}$}. The strength of a validation protocol is affected by the limited number $\mathcal{S}$ of collected events, as compared to the total number of distinct $n$-photon output events. While this limitation is not relevant for small-scale implementations, due to ($i$) the then low dimension of Hilbert space, ($ii$) a high level of control and ($iii$) reduced losses, it represents one of the main bottlenecks for the actually targeted large-scale instances \cite{Brugger18MSc}. It is thus desirable to assess the robustness and the resilience of a protocol under such incomplete-sampling effects, to quantify the impact of always strictly finite experimental resources on the actual applicability range of the protocol. We therefore propose to define a (minimal) threshold sample size $\mathcal{S}$ which must be available for validation. Given a set of $\mathcal{S}$ events, a validation protocol must be capable to give a reliable answer within a certain confidence level.
\item \textit{Available sampling time $\mathcal{T}$}. While the sampling rate is nearly constant for current quantum and classical approaches \cite{Clifford18}, \textit{de facto} making the time $\mathcal{T}$ not relevant, it cannot be excluded that future algorithms may process data and output all events at once. The very quality of the simulation, i.e. the similarity to quantum Boson Sampling in a given metric, could also improve with processing time \cite{Neville17, adversarial-ML}. Ultimately, $\mathcal{T}$ must be treated as an independent parameter with respect to $\mathcal{S}$, while at the same time it should be adapted to the sample size required for a reliable validation.
\item \textit{Unitary $\mathcal{U}$}. Unitary evolutions should be drawn Haar-randomly by a third agent, at the start of the competition to avoid any preprocessing. This agent, the \emph{validator} ($\mathcal{V}$), uses specific validation protocols to decide whether a sample is compatible with quantum operation.
\end{enumerate}

In the thus defined setting, a data set is said validated according to the following rule (Fig. \ref{Fig1}a): \newline

\noindent  \textit{Boson Sampling is validated if, collecting $\mathcal{S}$ events in time $\mathcal{T}$ from some random unitary $\mathcal{U}$, it is accepted by all selected validators $\mathcal{V}$.}\newline

Given a unitary and a set of validation protocols, we are then left with the choice of $\mathcal{S}$ and $\mathcal{T}$, which need be plausible for technological standards. Demanding to sample $\mathcal{S}$ events in time $\mathcal{T}$, these thresholds in fact limit the size of the problem ($n$,$m$) for an experimental implementation.
As for the time $\mathcal{T}$, one possibility, feasible for quantum experiments, could be for instance one hour. Within this time, a quantum device will probably output events at a nearly constant rate, while a classical computer can output them at any rate allowed by its clock cycle time.
The choice of the sample size $\mathcal{S}$ is instead more intricate, since a value too high collides with the limited $\mathcal{T}$, while a value too low implies an unreliable validation $\mathcal{V}$.
With these or further considerations \cite{jointV}, classical and quantum samplers should agree upon a combination of ($n$, $m$, $\mathcal{S}$, $\mathcal{T}$) that allows them to validate their operation.

\section{Validation with finite sample size}

\label{sec_partII}

In this section, we investigate a convenient approach to validation that distinguishes between two regimes: until $n \sim 30$ (Sec. \ref{subsec:bayesian}) and from $n \sim 30$ (Sec. \ref{subsec:benchmark}). In each section, we will first summarize the main ideas behind their operation. Then, we will discuss their performance for various ($n$, $m$), highlighting strengths and limitations, by numerically simulating experiments with finite sample size and distinguishable or indistinguishable photons.


\subsection{Bayesian tests for small-scale experiments}

\label{subsec:bayesian}

The Bayesian approach to Boson Sampling validation ($\mathcal{V}_B$), introduced in Ref. \cite{Bentivegna14}, aims to identify the most likely between two alternative hypotheses, which model the multi-photon states under consideration. 
In particular, $\mathcal{V}_B$ tests the Boson Sampling hypothesis ($H_{Q}$), which assumes fully indistinguishable $n$-photon states, against an alternative hypothesis ($H_{A}$) for the source that produces the measurement outcomes $\{x\}$. Equal probabilities are assigned to the two hypotheses prior to the experiment. Let us denote with $p_{Q}(x_k)$ $\left( p_A(x_k) \right)$ the scattering probability associated with the output state $x_k$ for $H_{Q}$ ($H_{A}$). The intuition is that, if $H_{Q}$ is most suitable to model the experiment, it is more likely to collect events for which  $p_{Q}(x_k)>p_{A}(x_k)$. The idea is made quantitative considering the confidence $ P(\{x\}|H_{hypo}) = \prod_{k=1}^\mathcal{S} p_{hypo}(x_k) $ we assign to each hypothesis. By applying Bayes' theorem, after $\mathcal{S}$ events we have
\begin{equation}
\frac{P(H_{Q}|\{x\})}{P(H_{A}|\{x\})}=\prod_{k=1}^\mathcal{S} \left( \frac{p_{Q}(x_k)}{p_A(x_k)}\right) = \chi_\mathcal{S},
\label{conf_Bayes}
\end{equation}
\noindent
and our confidence in the hypothesis $H_{Q}$ becomes $P(H_{Q}|\{x\}) = \frac{\chi_\mathcal{S}}{1+\chi_\mathcal{S}} $.

\begin{figure}[t]
\includegraphics[width=\linewidth]{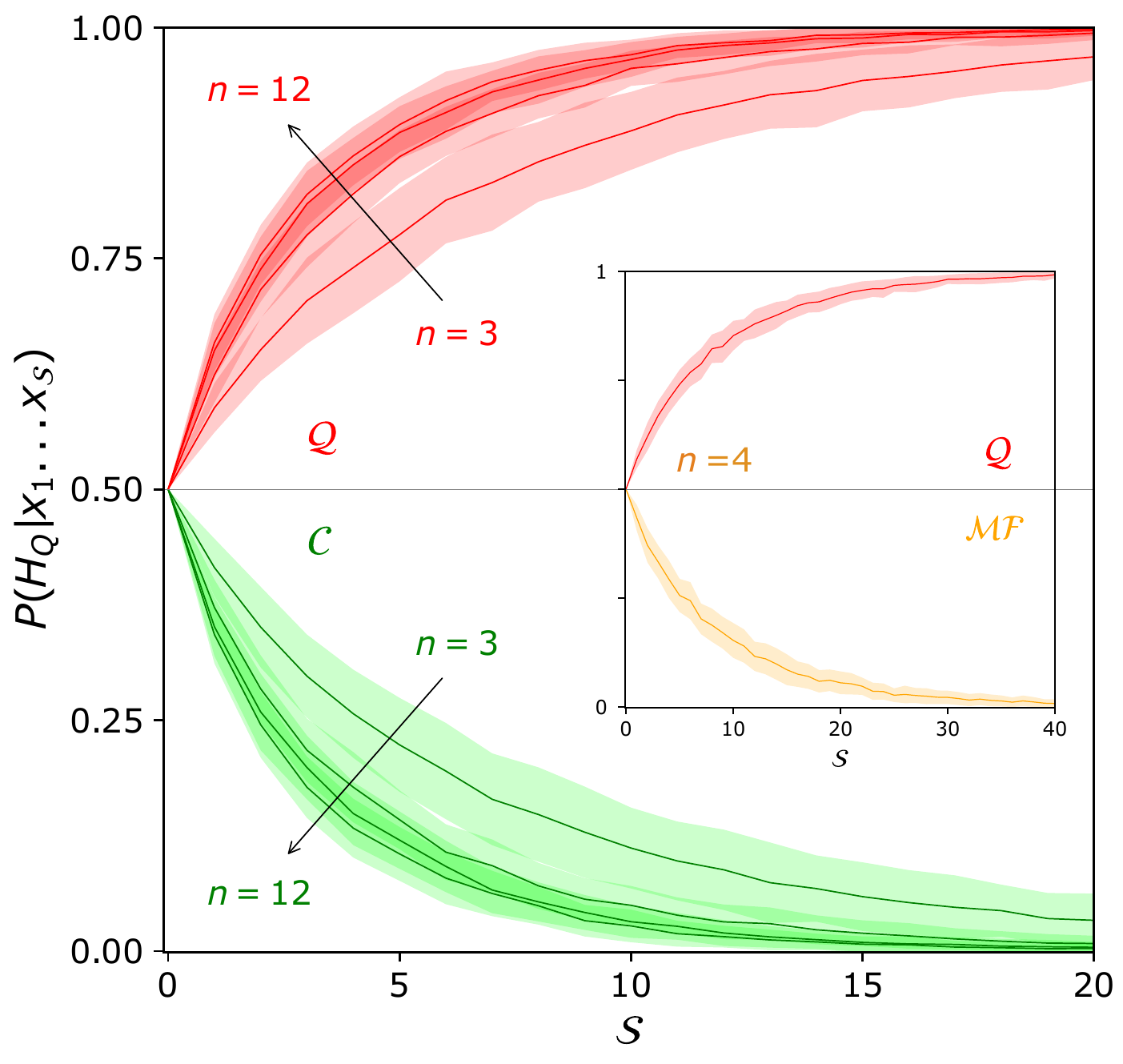}
\caption{
Confidence $P(H_{Q}|\{x\})$ of the Bayesian test to accept, as a correct Boson Sampling experiment, events that are sampled using distinguishable ($\mathcal{C}$, green) \cite{Aaronson14} and indistinguishable ($\mathcal{Q}$, red) \cite{Clifford18} $n$-photon states from $m$-mode interferometers. Note how curves become steeper for increasing $n$ ($n=3, 6, 9, 12$ and $m=n^2$), making the test progressively more sample-efficient. Inset: Bayesian protocol applied to test $\mathcal{Q}$ against the Mean-Field sampler ($\mathcal{M}\mathcal{F}$, orange) \cite{Tichy14} for $n=4$ photons and $m=n^2$. Curves are obtained by numerically sampling $10^4$ $n$-photon events, averaging over 50 random reshuffling of these events and over 100 different Haar-random unitary transformations (shaded regions: one standard deviation).
}
\label{Fig2}
\end{figure}

This test requires the evaluation of permanents of $n \times n$ scattering matrices for $p_{Q}(x_k)$ \cite{Valiant79, Wu16}, which sets an upper limit to the number of photons that can be studied in practical applications \cite{Wang16beating, Wang17, Wang16ten-photon, He17, Neville17, Viggianiello17tvd, Paesani19, Wang1920photons}. 
Indeed, it is foreseeable that real-time monitoring or feedback-loop stabilization of quantum optics experiments will only have access to portable platforms with limited computational power. However, an interesting advantage of this validation protocol is its broad versatility, due to the absence of assumptions on the alternative distributions. Importantly, when applied to validate Boson Sampling with distinguishable photons, it requires very few measurements ($\mathcal{S} \sim 20)$ for a reliable assessment. In Fig. \ref{Fig2}, for instance, we numerically investigate its application as a function of sample size, extending previous simulations from $n=3$ \cite{Bentivegna14} to $n=(3,6,9,12)$ and $m=n^2$.
%
%
Data for distinguishable ($H_{C}$) and indistinguishable ($H_{Q}$) photons were generated using exact algorithms, respectively by Aaronson and Arkhipov \cite{Aaronson14} and by Clifford and Clifford \cite{Clifford18}. The analysis shows how the validation protocol becomes even more effective for increasing $n$, being it able to output a reliable verdict after only $\sim 20$ events. However, as mentioned, its power comes at the cost of being computationally inefficient in $n$. Also, it is not possible to preprocess $\mathcal{V}_B$ and store information for successive re-use, since its confidence depends on the specific $\mathcal{U}$ and sampled events, according to $p_{Q}(x_k)$.
Hence, in the regime $n \sim 25-35$ \cite{Wang1920photons, Neville17} it becomes rapidly harder to perform a validation in real time. Eventually, since classical supercomputers cannot assist quantum experiments in everyday applications, $\mathcal{V}_B$ becomes prohibitive from $n\sim35$.


\subsection{Statistical benchmark for large-scale experiments}

\label{subsec:benchmark}

\begin{figure*}[ht!]
\includegraphics[width=\linewidth]{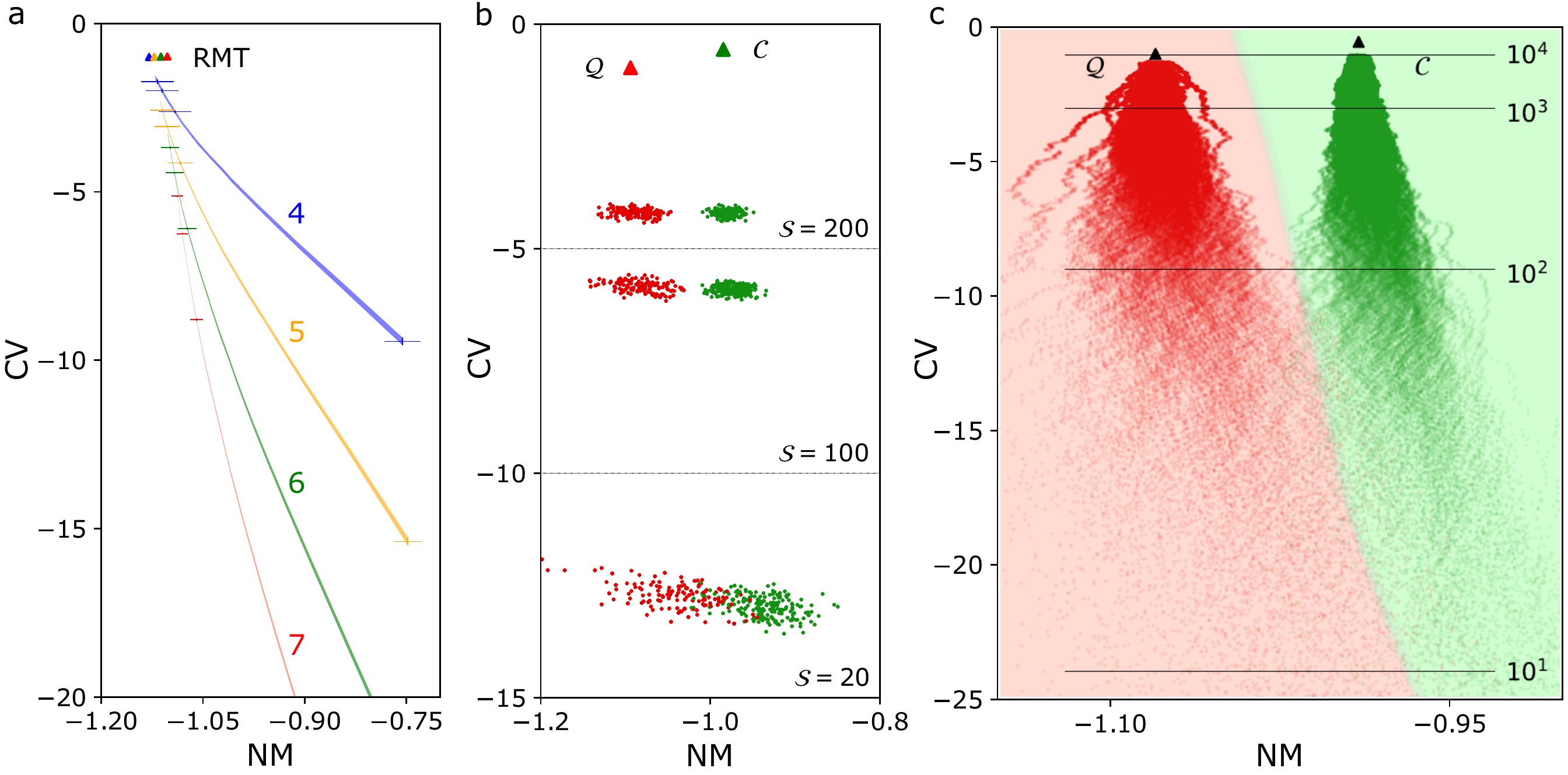}
\caption{
a) Numerically simulated evolution of $C$-datasets in the NM-CV plane for an increasing sample size $\mathcal{S}$. Boson Sampling with $n=(4,5,6,7)$ indistinguishable photons in $m=n^2$ modes, for $\mathcal{S}=2,3,4,...,100$. For large $\mathcal{S}$, curves converge to the points (pyramids) predicted by random matrix theory (RMT) \cite{Walschaers16}. Points are averaged over $100$ Haar-random unitaries, while error bars are displayed every 20 additional events.
b) Validation via statistical benchmark of numerically generated Boson Sampling data with indistinguishable photons (red) against ones with distinguishable photons ($\mathcal{C}$, green), for 100 different unitary transformations, $n$=8 photons, $m$=64 modes and $\mathcal{S}=20,100,200$ events. Contour plots describe the confidence of (three different instances of) neural network binary classifiers, trained at different $\mathcal{S}$, from green (label: $\mathcal{C}$) to red (label: $\mathcal{Q}$). Red and green pyramids identify the random matrix prediction from Ref. \cite{Walschaers16}, for $S\rightarrow\infty$. Note how the clouds of points shrink for increasing $\mathcal{S}$, making the classification of experimental points progressively more reliable. c) Same analysis as in (b), for 200 different unitary transformations, $n$=10 photons, $m$=100 modes and $\mathcal{S}=1, 2, 3, ... 10^4$ events. The classifier is now trained on data from all $\mathcal{S}$. Numbers on the right indicate, approximately, the sample size at the corresponding height. Once trained, this classifier can be deployed to validate any other Haar-random experiment with the same ($n$, $m$) and any $\mathcal{S}$.
}
\label{Fig3}
\end{figure*}

\begin{figure}[h]
\includegraphics[width=\linewidth]{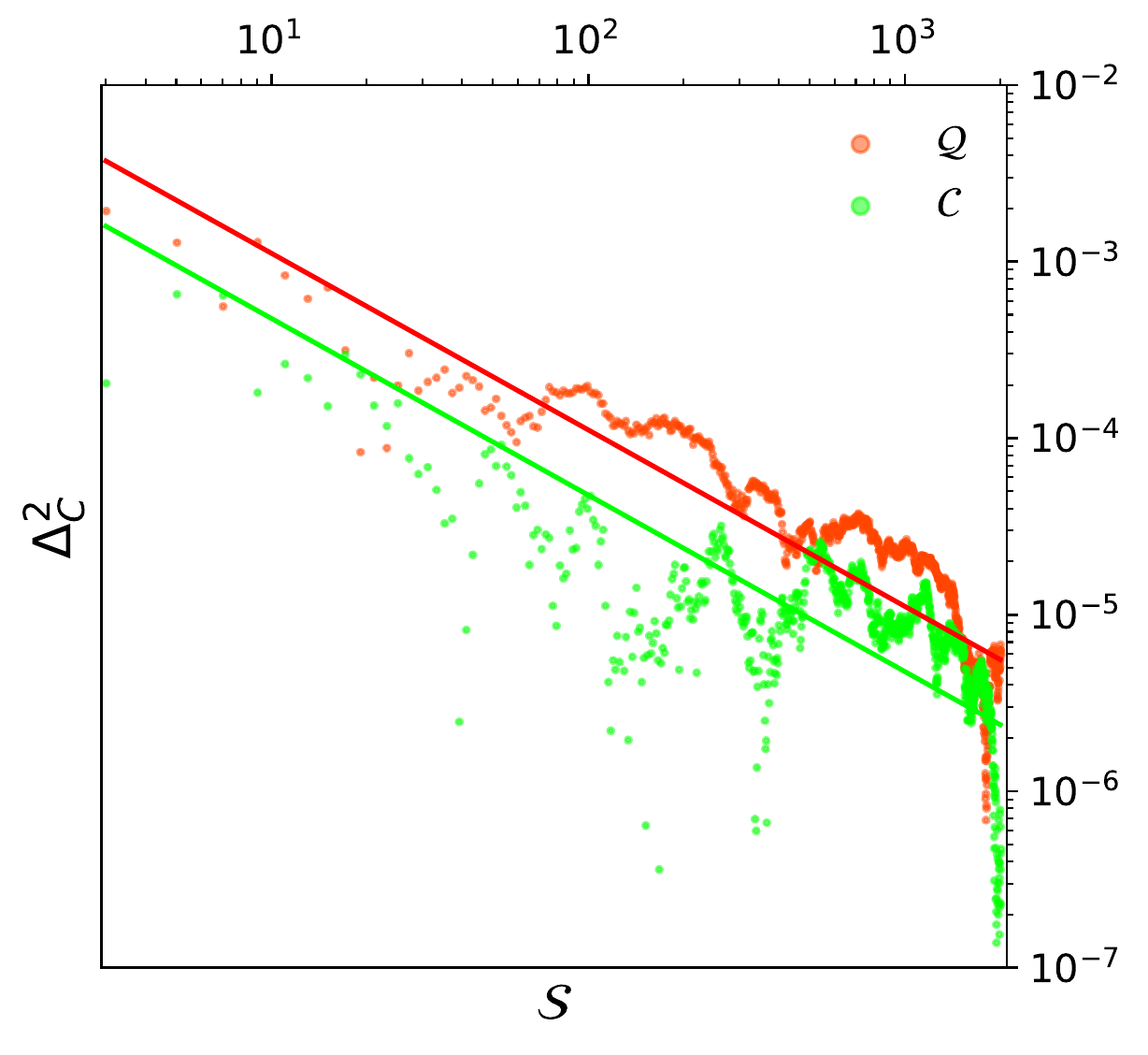}
\caption{
Log-Log plot of the deviation $\Delta C^2_\mathcal{S} = | \mathds{E}_{U}[\mathds{E}_{X} ({\widetilde{C}^2_{ij}})] - \mathds{E}_{U}[{C^2_{ij}}] | = | \mathds{E}_{U}[\sigma_{ij}^2]/\mathcal{S} | $ from Eq. \ref{eq.meanCorr2} as a function of the sample size $\mathcal{S}$. Data numerically generated to mimic experiments with $n=4$ photons in $m=16$ modes (green: distinguishable photons \cite{Aaronson14}; red: indistinguishable photons \cite{Clifford18}). Averages are carried out over 500 Haar-random unitaries $U$ and 500 different samples of size $\mathcal{S}$ (number of events) from each unitary, with fixed input state (1,1,1,1,0,...,0). The linear fits to the different data sets exhibit the expected scaling  $\propto \mathcal{S}^{-1}$.
}
\label{Fig4}
\end{figure}

In the previous section we described how the Bayesian test is effective in validating small- and mid-scale experiments with very few measurement events. However, the evaluation of permanents hinders its application for large $n$, be it due to too large scattering matrices or to the need for speed in real-time evaluations. To overcome this limitation, further validation protocols have been proposed in the last few years, to find a convenient compromise between predictive power and physical resources. All these approaches have their own strengths and limitations, and tackle the problem from different angles \cite{Walschaers19}, e.g. using suppression laws \cite{Tichy10, Tichy14, Crespi15, Dittel17, Dittel18}, machine learning \cite{Agresti19, Flamini19tsne} or statistical properties related to multi-particle interference \cite{Walschaers16}. In this section we will focus on the latter protocol, which arguably represents the most promising solution for the reasons we outline below.

\textit{Statistical benchmark with finite sample size.} Validation based on the statistical benchmark ($\mathcal{V}_{S}$) looks at statistical features of the \emph{C-dataset}, the set of two-mode correlators
\begin{equation}
C_{ij}=\langle \hat{n}_i  \hat{n}_j \rangle - \langle \hat{n}_i \rangle \langle \hat{n}_j \rangle
\label{C-dataset}
\end{equation}
\noindent where $(i,j)$ are distinct output ports and $\hat{n}_i$ is the bosonic number operator.
Two statistical features that are effective to discriminate states with indistinguishable and distinguishable photons are its normalized mean NM (the mean divided by $n/m^2$) and its coefficient of variation CV (the standard deviation divided by the mean). For any unitary transformation and input state we can retrieve a point in the plane (NM, CV), where alternative models tend to cluster in separate clouds located via random matrix theory (Fig. \ref{Fig3}a) \cite{Walschaers16}.
Validation based on $\mathcal{V}_{S}$ would then consist in (i) collecting a suitable number $\mathcal{S}$ of events, (ii) evaluating the experimental point (NM, CV) associated to the $C_{ij}$ and (iii) identifying the cluster that the point is assigned to. For $\mathcal{S}$ sufficiently large, the point will be attributable with large confidence to only one of the models, thus ruling out the others (Fig. \ref{Fig3}b). 

$\mathcal{V}_{S}$ represents the state of the art for validation protocols that do not require the evaluation of permanents. Indeed, this approach has several advantages \cite{Giordani18}:
($a$) it is computationally efficient (one only needs to compute two-point correlators), ($b$) it can reveal deviations from the expected behaviour (manifest in the NM-CV plane), ($c$) it makes more reliable predictions for larger $n$ (clouds become more separate), ($d$) it is sample-efficient (clouds separate relatively early, after few measurements events). However, despite points ($c$, $d$) above, in actual conditions the experimental point is not always easy to validate. In fact, as mentioned in point ($b$), hardware imperfections and partial distinguishability make the point move away from the average route shown in Fig. \ref{Fig3}a. These issues can be addressed and mitigated by numerically generating, for a fixed sample size $\mathcal{S}$, clouds from unitary transformations that take these aspects into account. As suggested in Ref. \cite{Giordani18}, and more closely investigated in Fig. \ref{Fig3}b,c, a convenient approach is to employ machine learning to assign experimental points to one of the two clouds, with a certain confidence level. Specifically, one can train a classifier with numerically generated data \cite{Aaronson14, Clifford18} for a certain ($n$, $m$, $\mathcal{S}$), that can even include error models, and then deploy it for all applications in that regime.  In this sense, $\mathcal{S}$ can be seen as the label of the model that can classify (validate) data for a given ($n$, $m$). This intuition can be extended to a classifier that is trained on data from multiple $\mathcal{S}$ (see Fig. \ref{Fig3}c), which is likely more practical. For a fixed $\mathcal{S}$, the computational resources to sample events from a distribution given by $n$ distinguishable (indistinguishable) photons scale polynomially \cite{Aaronson14} (exponentially \cite{Clifford18}) in $n$. However, once trained, this classifier can be considered as an off-the-shelf tool that is readily applicable to validate multi-photon interference with no additional computational overhead, which is ideal for large-size experiments. In Sec. \ref{sec:Appendix_AdaBoost}, we also discuss how such a classifier can even be combined with other protocols, which search the data for different distinctive structures, to boost its accuracy.

\textit{Finite-size effects in validation protocols.}
So far, we qualitatively discussed the role of a limited sample size for the validation of multi-photon quantum interference. To provide a more quantitative analysis of finite-size effects for the task of validation, and in particular for $\mathcal{V}_{S}$, in the following we study the scaling of the parameters involved in the above validation protocol with $\mathcal{S}$. The goal of this section is to elaborate on a standard test which should be implemented in all validation protocols, to guarantee their experimental feasibility. 

Let us start by considering a fixed unitary circuit $U$, for which we calculate the correlators $C_{ij}$ from Eq. (\ref{C-dataset}). Such evaluation in principle assumes the possibility to collect an arbitrary number of measurement events.  In practical applications, however, sample sizes will always be limited. Hence, finite-size effects play a role in the estimation of the above correlators. According to the central limit theorem, the correlator retrieved from the experimental data can be represented as $\widetilde{C}_{ij} = C_{ij} + X_{ij}$, where $X_{ij}$ is a random number normally distributed with zero mean and variance $\sigma_{ij}^2\, \mathcal{S}^{-1}$. The $\sigma_{ij}^2$ depend on the unitary evolution $U$ and should either be evaluated from the data or be estimated using random matrix theory. Now, to infer, from noisy C-datasets \cite{Walschaers16}, the centre of the cloud of points in the NM-CV plane, we need to average not only over the Haar measure, but also over $X_{ij}$.

Consequently, we have to assess the impact of finite-size effects on the estimate of the moments (NM, CV). First, since the noise induced by the finite sample size averages out, namely $\mathds{E}_{X} (\widetilde{C}_{ij}) = {C}_{ij}$, we have that $\widetilde{NM} = NM$.
The estimation of CV is a bit more subtle because we need to evaluate the mean of ${\widetilde{C}_{ij}}^2$. Since $\mathds{E}_{X} ({\widetilde{C}_{ij}}^2) = {C^2_{ij}} + \sigma_{ij}^2 \, \mathcal{S}^{-1}$, then

\begin{equation}
\mathds{E}_{U}[\mathds{E}_{X} ({\widetilde{C}^2_{ij}})] = \mathds{E}_{U}[{C^2_{ij}}] + \frac{\mathds{E}_{U}[\sigma_{ij}^2]}{\mathcal{S}}
\label{eq.meanCorr2}
\end{equation}

\noindent and, hence, $| \widetilde{CV} | > | CV |$. Note that $\mathds{E}_{U}[\mathds{E}_{X} ({\widetilde{C}^2_{ij}})]$ and $\mathds{E}_{X}[\mathds{E}_{U} ({\widetilde{C}_{ij}}^2)]$ cannot be easily compared, since the latter involves averaging the distribution of $X_{ij}$ over the unitary group. However, using the properties of the normal distribution under convex combinations, we can deduce that both orders of averaging yield approximately the same result (and the same scaling in $\mathcal{S}$), in particular once $\mathcal{S}$ is large and the distribution is concentrated close to its mean. Numerical simulations for $3\le n \le 15$ and $m=n^2$ indeed confirm its validity (Fig. \ref{Fig4}).
Specifically, we observe that, upon averaging over different Haar-random unitaries with $\mathcal{S}$ events per realization, the deviation of the experimentally-measured ${\widetilde{C}_{ij}}^2$ from the analytically predicted values decreases as fast as $1/\mathcal{S}$. Hence, their estimation from finite-size data sets shows no exponential overhead that would hinder a practical application of the validation protocol.


\section{Discussion} 

Validation of multi-photon quantum interference is expected to play an increasing role as the dimensionality of photonic applications increases, both in the number of photons and modes. To this end, and as notably emphasized by the race towards quantum advantage via Boson Sampling, it is necessary to define a set of requirements for a validation protocol to be meaningful. Ultimately, these requirements should allow to establish strong experimental evidence of quantum advantage that is accepted by the community within a jointly agreed framework.

In the present work, we implement such a program and describe a set of critical points that experimenters will need to agree upon in order to validate the operation of a quantum device. 
With the goal of building a solid framework for validation, we then discuss a practical approach that applies the most suitable state-of-the-art protocols in realistic scenarios. We report numerical analyses on the application of two key validation protocols, the Bayesian hypothesis testing and the statistical benchmark, with finite-size data, providing compelling evidence in support of this approach.

A clear and illustrative example for the above considerations is provided in Section \ref{Appendix_MF_Markov}, where we numerically studied the competition between a recent classical simulation algorithm and the statistical benchmark, respectively to counterfeit and to validate Boson Sampling, while they process an increasing number of measured output events. The analysis quantifies the general intuition that there must be a trade-off between speed and quality in approximate simulations of Boson Sampling. We also provide a formal analysis on the performance of the validation protocol with finite-size samples, showing that the estimation of relevant quantities converges fast to the predicted values. We expect that similar features will be crucial for larger-scale demonstrations and, as such, a key prerequisite to be investigated in all validation protocols.

Finally, in Section \ref{sec:Appendix_AdaBoost} we introduce a novel approach to validation that can bring together the strengths of multiple protocols. This approach uses a meta-algorithm (AdaBoost) to combine protocols based on machine learning into a single validator with boosted accuracy. This strategy becomes more advantageous for a larger number of such protocols with comparable performance, as well as with very noisy data.


\section*{Acknowledgements} 
This work was supported by the ERC Advanced Grant CAPABLE (Composite integrated photonic platform by femtosecond laser micromachining; grant agreement no. 742745); by the QuantERA ERA-NET Cofund in Quantum Technologies Project HiPhoP (High dimensional quantum Photonic Platform, Project ID 731473) and by project PRIN 2017 "Taming complexity via QUantum Strategies a Hybrid Integrated Photonic approach" (QUSHIP) Id. 2017SRNBRK. 
A.B. acknowledges support by the Georg H. Endress foundation.
M.W. is funded through Research Fellowship WA 3969/2-1 of the German Research Foundation (DFG).
This project has received funding from the European Union’s Horizon 2020 research and innovation programme under the Marie Skłodowska-Curie grant agreement No 801110 and the Austrian Federal Ministry of Education, Science and Research (BMBWF). It reflects only the author’s view and the Agency is not responsible for any use that may be made of the information it contains.

\section{Appendix} 

\subsection{Classical simulation and the role of sample size}

\label{Appendix_MF_Markov}

\begin{figure}[h]
\includegraphics[width=0.95\linewidth]{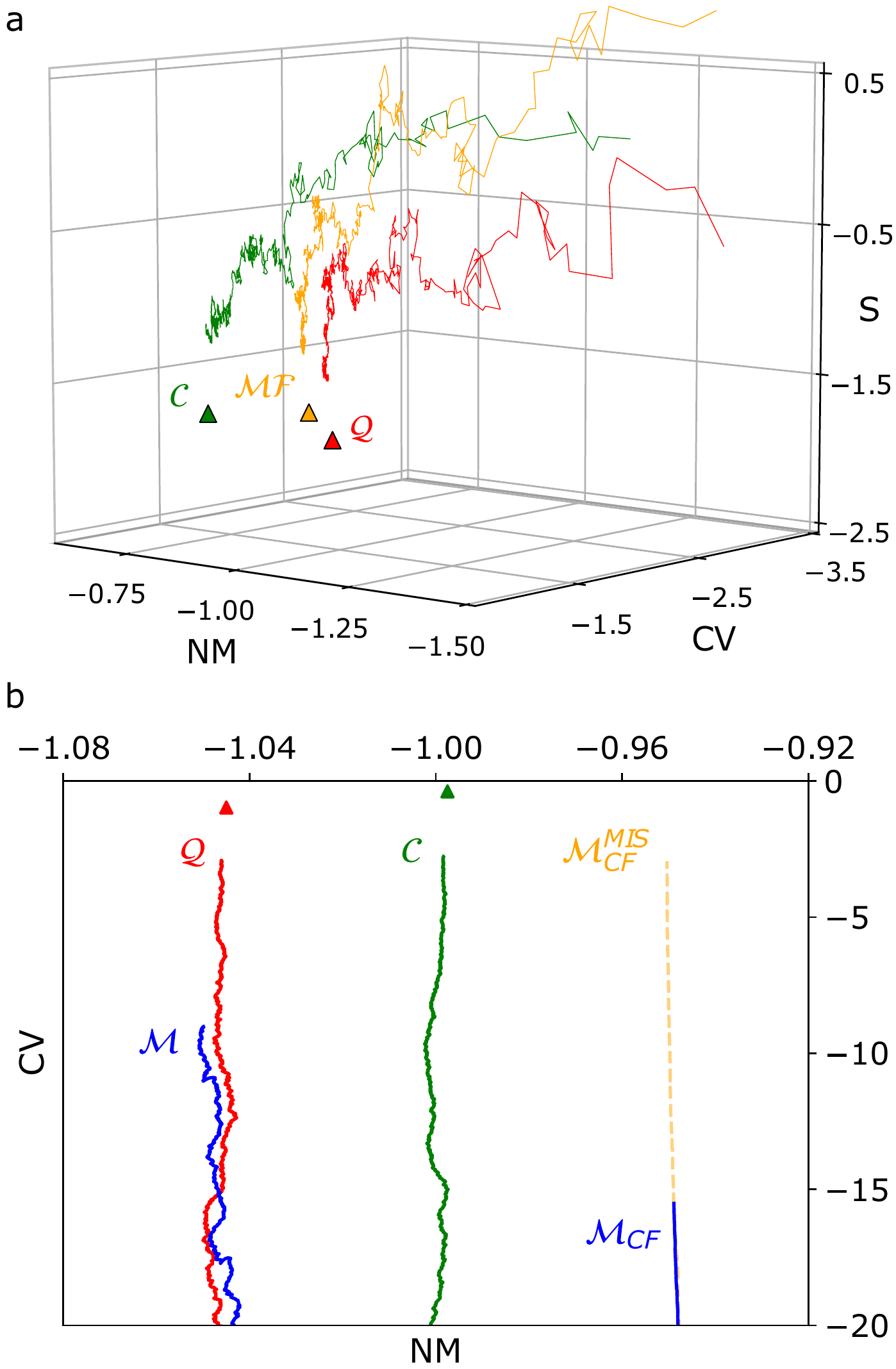}
\caption{
Numerically simulated evolution of C-datasets in the NM-CV-S space (a) and in the NM-CV plane (b) for an increasing sample size $\mathcal{S}$. 
a) Boson Sampling with $n=4$ indistinguishable photons ($\mathcal{Q}$, red) or distinguishable photons ($\mathcal{C}$, green) and Mean-Field sampler ($\mathcal{MF}$, orange) \cite{Tichy14} in an $m=16$-mode Haar-random transformation, for $\mathcal{S}= 10^4$ events. Pyramids identify the random matrix prediction for $S\rightarrow\infty$ \cite{Walschaers16}. 
b) Boson Sampling with indistinguishable or distinguishable photons and Metropolized independent sampling (MIS), with collision events ($\mathcal{M}$) or without ($\mathcal{M}^{\textup{\tiny MIS}}_{\textup{\tiny CF}}$, extracted from data in Ref. \cite{Neville17}; $\mathcal{M}_{\textup{\tiny CF}}$, data subset of $\mathcal{M}$) for $n$=20 photons, $m$=400 modes and up to $\mathcal{S}=2 \times 10^4$ events. Curves without collision events (which can be resolved under stronger zoom) have a smoother evolution due to reduced fluctuations in the $C$-dataset.
Note that the statistical benchmark captures the presence of collision events (in $\mathcal{Q}, \mathcal{C}, \mathcal{M}$), which have an impact on the statistics since the protocol probes two-particle processes.
}
\label{FigA1}
\end{figure}

To shed some light on the critical aspects of validation, and as a benchmark of the state of the art in this context, we now provide a qualitative analysis inspired by the Metropolized independent sampling ($\mathcal{M}$), a recent algorithm to classically simulate Boson Sampling \cite{Neville17}. 
The idea behind $\mathcal{M}$ is reminiscent of the Mean-Field sampler ($\mathcal{MF}$) \cite{Tichy14}, an adversarial classical algorithm that was capable to hack one of the first validation protocols \cite{Carolan14} using limited classical resources. In the race towards quantum computational supremacy, the introduction of $\mathcal{MF}$ has prompted the development of more sophisticated techniques to tackle classical simulations. For instance, besides the Bayesian test (see inset in Fig. \ref{Fig2}), also the statistical benchmark is highly effective to validate Boson Sampling against $\mathcal{MF}$ (see Fig. \ref{FigA1}a). For our scope, the key difference between the two algorithms is that, while for $\mathcal{MF}$ the quality of the simulation does not really change over time, $\mathcal{M}$ samples from a distribution that gets closer to $\mathcal{Q}$ the more events are evaluated (i.e. for a larger $\mathcal{S}$).

\begin{figure}[h!]
\includegraphics[width=0.9\linewidth]{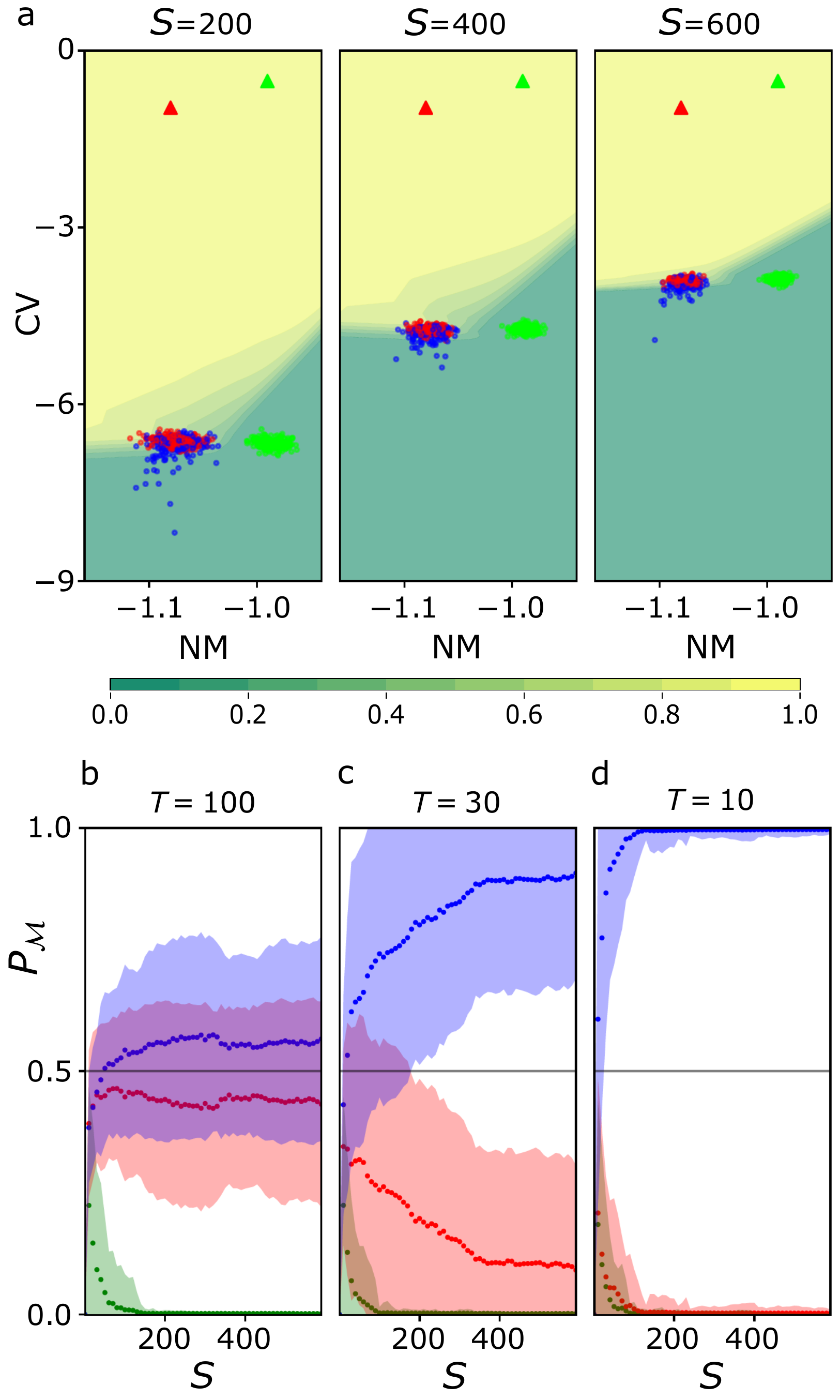}
\caption{
a) Validation via statistical benchmark \cite{Walschaers16} of Boson Sampling with indistinguishable photons ($\mathcal{Q}$, red) against one with distinguishable photons ($\mathcal{C}$, green) and Metropolized independent sampling \cite{Neville17} ($\mathcal{M}$, blue) with thinning $T=100$ and burn-in $B=0$, for 200 simulated experiments with different Haar-random unitary transformations, $n$=10 photons, $m$=100 modes and $\mathcal{S}=200,400,600$ events. Contour plots describe the confidence of a neural network classifier, from green (low) to yellow (high), in labeling a point as $\mathcal{Q}$. Red and green pyramids identify the random matrix prediction from Ref. \cite{Walschaers16}, for $S\rightarrow\infty$. 
(b,c,d) Confidence $P_\mathcal{M}$ of the same classifier in labeling (NM,CV) points generated by $\mathcal{M}$ as $\mathcal{M}$ (blue), $\mathcal{Q}$ (red) or $\mathcal{C}$ (green), for simulated experiments with $T=100$ (b) from (a), $T=30$ (c) and $T=10$ (d). Values are averaged over all (NM,CV) points generated by $\mathcal{M}$, while shaded regions correspond to one standard deviation. Notice that in (b), with strong thinning, there still is a difference between $\mathcal{Q}$ and $\mathcal{M}$ data, though not significant due to larger fluctuations. Plots highlight the speed vs. quality trade-off in classical simulations of Boson Sampling. See the main text for a step-by-step description of this analysis.
}
\label{FigA2}
\end{figure}

The goal of $\mathcal{M}$ is to generate a sequence of $n$-photon events $\{e_i\}$ from a Markov chain that mimics the statistics of an ideal Boson Sampling experiment. Given a sampled event $e_{i}$, a new candidate event $e_{i+1}$ is efficiently picked according to the probability distribution of distinguishable photons $p_D$, and accepted with probability

\begin{equation}
P(e_{i} \rightarrow e_{i+1}) = \textup{min}\left(1,\; \frac{p_I(e_{i+1}) \; p_D(e_{i})}{p_D(e_{i+1}) \; p_I(e_{i})} \right) 
\label{markov}
\end{equation}

\noindent where $p_I(e_{i})$ is the output probability corresponding to event $e_{i}$ for indistinguishable photons. While the approach remains computationally hard, since it requires the evaluation of permanents \cite{Scheel08,Valiant79}, the advantage is that only a limited number of them needs to be evaluated to output a new event, rather than the full distribution as in a brute-force approach. Ultimately, after a certain number of steps in the chain, $\mathcal{M}$ is guaranteed to sample close to the ideal Boson Sampling distribution $p_I$ \cite{Brooks98}. Hence, not only does the sample size $\mathcal{S}$ play a key role to improve the reliability of validation protocols, as shown in Sec. \ref{sec_partII}, but it can be crucial also to increase the quality of the outcome of a classical simulation. This is a relevant point to keep in mind, even though $\mathcal{M}$ has since been surpassed by an algorithm that is both provably faster and exact \cite{Clifford18}. In fact, in future, novel classical algorithms might be developed \cite{Liu19} that depend on $\mathcal{S}$ more efficiently.

The aim of our present analysis is to investigate the role of the sample size in a validation of the samples generated by $\mathcal{M}$, via $\mathcal{V}_S$. Indeed, a crucial issue in a hypothetical competition between $\mathcal{M}$ and $\mathcal{V}_S$ concerns the number of events $\mathcal{S}$ available to accept or reject a data set. While larger sets provide deeper information to $\mathcal{V}_S$ to identify fingerprints of quantum interference, on the other hand $\mathcal{M}$ approaches the target distribution $p_I$ as more steps are made along the chain. However, in order to output a large number of events in time $\mathcal{T}$, $\mathcal{M}$ requires physical and computational resources that set a limit to the tractable dimension of the problem. We are then interested in the intermediate regime, the one relevant for experiments, to determine whether convergence is reached fast enough to mislead $\mathcal{V}_S$. In the specific case of $\mathcal{M}$, we then need to look at the scaling in $n$ of its hyper-parameters: \textit{burn-in} (the number $B_n$ of events to be discarded at the beginning of the chain) and \textit{thinning} (the number $T_n$ of steps to skip to reduce correlations between successive events). Eventually, the time required to classically simulate Boson Sampling will scale as $\mathcal{T} = \tau_{p} \left(B_n + \mathcal{S} \; T_n  \right)$, where $ \tau_{p} $ is the time to evaluate a single scattering amplitude according to Eq. (\ref{markov}). Considering the estimate provided by the supercomputer Tianhe-2 \cite{Wu16}, and for fixed ($\mathcal{T}$, $\mathcal{S}$), we find the constraint
$B_n = \alpha \,n^{-2} \,2^{-n} \;\mathcal{T}  -  \mathcal{S} \; T_n $
where $\alpha \sim c^{0.8782} \,10^{11}$ and $c$ is the number of processing nodes. If we assume $T_n=100$ \cite{Neville17} for all $n$ and $\mathcal{V}$, we get an estimate of the maximum $B_n$ allowed by ($\mathcal{T}$, $\mathcal{S}$). The key issue is that this estimate does not guarantee that $\mathcal{M}$ achieves the target distribution fast enough, since $B_n$ decreases (exponentially) in $n$. Moreover, the minimum $B_n$ is expected to increase with $n$, since on average the Markov chain needs to explore more states before picking a good one. 

To better clarify the above considerations, we simulate a competition between $\mathcal{M}$ and $\mathcal{V}_{S}$ for $n=10$ photons in $m=100$ modes on Fig. \ref{FigA2}.
Data for distinguishable and indistinguishable photons were generated with exact algorithms, respectively by Aaronson and Arkhipov \cite{Aaronson14} and by Clifford and Clifford \cite{Clifford18}.
The analysis proceeds through five main steps: 1) randomly pick a unitary transformation $\mathcal{U}$ according to the Haar measure; 2) simulate the generation of $\mathcal{S}$ $n$-particle output events; 3) extract the $C$-dataset from these $\mathcal{S}$ events; 4) evaluate the corresponding (NM, CV) point and plot it in Fig. \ref{FigA2}a; 5) repeat steps 1-4 200 times, to simulate as many different experiments. Upon completion, evaluate average and variance of $P_\mathcal{M}$ and plot them in Fig. \ref{FigA2}b. With this analysis,
we get a quantitative intuition on how the confidence of a validation changes with $\mathcal{S}$, as does the quality of the classical simulation. Similar behaviour is found also for other choices of $n$ and $m$. In particular, we observe how a stronger thinning (up to $T_{10}=100$, as in Ref. \cite{Neville17}) is reflected in the quality of the simulation, where $\mathcal{M}$ behaves very similar to the ideal Boson Sampler for small as well as for large sample sizes. Conversely, a faster $\mathcal{M}$ that trades quality for speed by computing fewer permanents ($T_{10}=10,30$) is more easily detectable by $\mathcal{V}_{S}$.
Constraints due to a speed vs. quality compromise (Fig. \ref{Fig3}b,c,d) define a generic scenario for a classical simulation which is run with a specific choice of $\mathcal{T}$ and $\mathcal{S}$.


\subsection{Combining and boosting validation protocols}

\label{sec:Appendix_AdaBoost}

So far, all validation protocols have always been applied separately and independently. Certainly, this fact shows the multifaceted nature of this line of research, where effective solutions have been developed using very different strategies. Yet, it also reflects its somewhat fragmented condition, since each protocol does not benefit from potential insights provided by the others. This limitation becomes relevant in realistic scenarios with noise and finite data sets, since each validation protocol suits some task better than the others, with different degrees of sample efficiency and resilience.

In this section, we present a novel, synergistic approach to validation, which aims at combining the strengths of these protocols to form a joint, enhanced validator. Specifically, we focus on validation protocols that make use of machine learning, and propose to combine them with a meta-algorithm (AdaBoost \cite{Freund97}) that attempts an adaptive boosting of their individual performance. The output of AdaBoost is a weighted sum of the predictions of these learning algorithms ('weak learners'), which are asked, sequentially, to pay more attention to the instances that were incorrectly classified by the previous learners. As long as the performance of each learner is slightly better than chance, the classifier resulting from AdaBoost provably converges to a better validation protocol.

\begin{figure}[t]
\includegraphics[width=\linewidth]{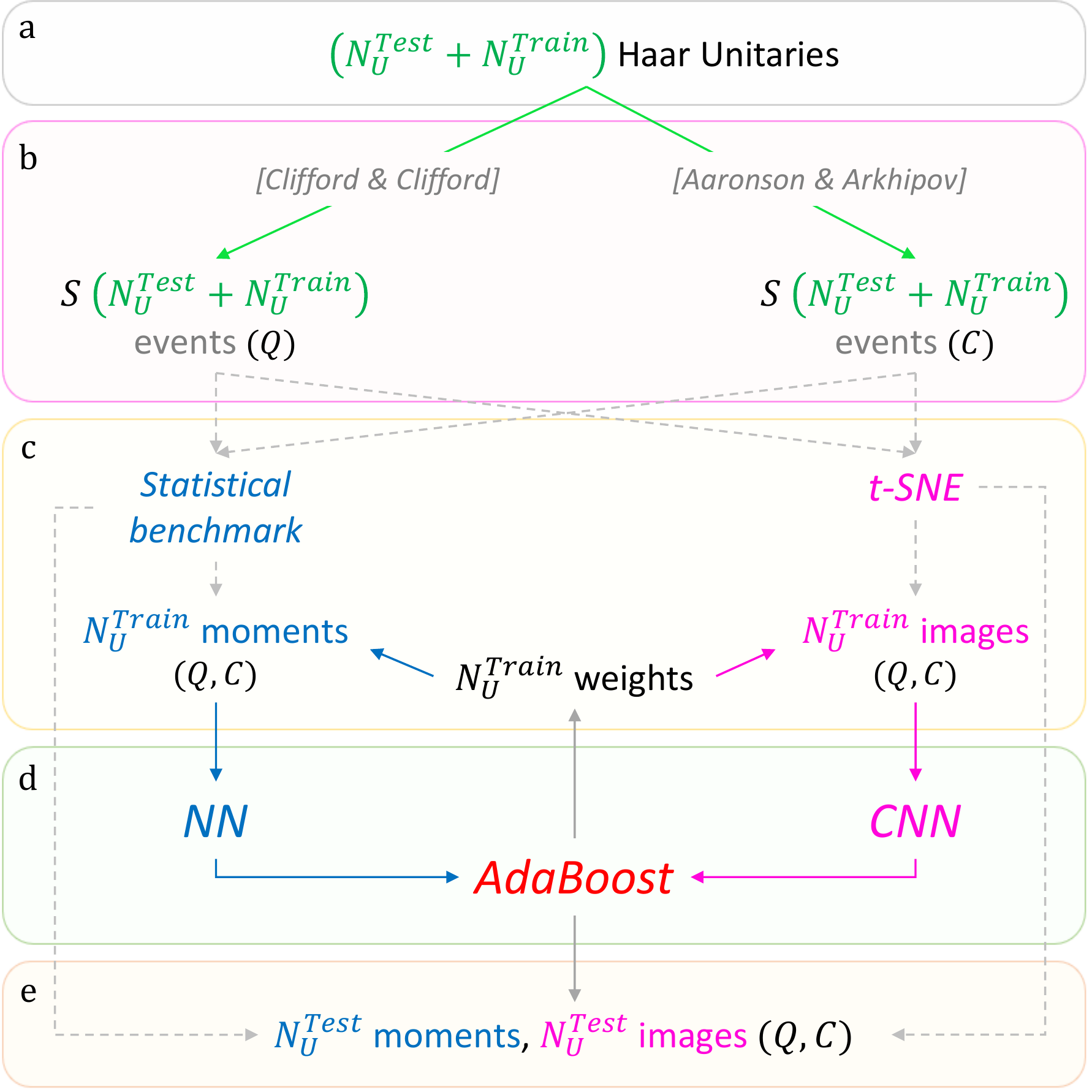}
\caption{
Machine learning techniques, such as AdaBoost \cite{Freund97}, can combine individual validation protocols to boost the overall accuracy. A schematic overview of the approach is shown in this figure. Experiments associated with $N_U^{Test}+N_U^{Train}$ Haar-random unitary transformations (a) are simulated using exact algorithms to numerically sample $\mathcal{S}$ events from quantum \cite{Clifford18} ($Q$) and classical \cite{Aaronson14} ($C$) Boson Sampling (b). c) $N_U^{Train}$ sets of $\mathcal{S}$ events for both $Q$ and $C$ are pre-processed by a collection of validation protocols. Here, we considered the statistical benchmark $\mathcal{V}_S$ \cite{Walschaers16} and the visual assessment $\mathcal{V}_C$ \cite{Flamini19tsne}, which produced $N_U^{Train}$ input data in the form of, respectively,  pairs of moments (NM, CV) and images (using t-SNE \cite{vanderMaaten08} for dimensionality reduction). d) Each protocol has its own classifier, in this case a neural network (NN) for $\mathcal{V}_S$ and a convolutional neural network (CNN) for $\mathcal{V}_V$, respectively. These classifiers are then applied sequentially on the same input data, iteratively adjusting their weights (AdaBoost) to focus on misclassifed data. e) The resulting, joint protocol has higher accuracy on test data than each individual classifier.
}
\label{FigA3}
\end{figure}

We numerically test this approach by combining two validation protocols that employ machine learning: the statistical benchmark $\mathcal{V}_S$ \cite{Walschaers16} (equipped with a simple neural network classifier trained on numerically generated data, as in Fig. \ref{Fig3}b,c) and the \emph{visual assessment} $\mathcal{V}_V$ \cite{Flamini19tsne}, which uses dimensionality reduction algorithms and convolutional neural networks. Here we do not consider the Bayesian approach, since, in its current formulation, it does not fit the framework of machine learning. A schematic description of our proof-of-concept analysis, which we carry out for $n=10$ and $m=100$, is shown in Fig. \ref{FigA3}.
%
%
Since $\mathcal{V}_S$ requires fewer events than $\mathcal{V}_V$ to validate ideal, noiseless experiments \cite{Aaronson14, Clifford18}, to perform this test we trained $\mathcal{V}_S$ on data sets with a tunable amount noise, purposely assembled to be hard to validate. To this end, samples ($\mathcal{S}=2\times10^3$) for 500 Haar-random unitary transformations were constructed by sampling with a certain probability $p$ (or $1-p$) from a Boson Sampler with fully indistinguishable (or distinguishable) photons. This probability $p$ was then varied in time, to simulate, for instance, a periodic drift in the synchronization of the input photons. As expected with these settings, we find that AdaBoost maintains the original accuracy of $\mathcal{V}_S$ and $\mathcal{V}_V$ when applied to, respectively, batches of $\mathcal{V}_S$ and $\mathcal{V}_V$ that are already highly accurate. This is mainly due to complexity of these classifiers, which are already strong learners and, hence, hard to enhance by AdaBoost. Analogous results are found with mixed batches of $\mathcal{V}_S$ and $\mathcal{V}_V$, for which AdaBoost returns a joint classifier that practically focuses on the most accurate one in the set. A different result is obtained, instead, by combining several weak $\mathcal{V}_V$, for which we purposely spoil the training of the convolutional neural network (accuracy $A \sim 51\%$ instead of $A \sim 98\%$) by reducing the number of training epochs. In this case, AdaBoost does in fact enhance the accuracy of $\mathcal{V}_V$ up to $A \sim 57\%$.

In future, we expect that this approach will prove useful in non-ideal conditions with experimental noise, where validation protocols do not operate in the ideal settings where they were conceived. Furthermore, the above analyses can show larger boosts if applied to actual experiments that involve structured (non-Haar-random) interferometers, for which protocols such as $\mathcal{V}_S$ and $\mathcal{V}_V$ can have lower accuracies and different behaviors. Finally, still in non-ideal settings, more favorable boosts can be obtained if new validation protocols are developed that are as sample-efficient as $\mathcal{V}_S$.

\section*{References}


\begin{thebibliography}{10}
\bibitem{jointV}
Given that the validators have a non-zero probability of false positives or negatives the criteria for validation need to be generalized. in this context a device is said to be validated if the probability that a valid device passes (or fails) the tests is greater by a fixed fraction than the probabilities that would be predicted by all available alternative hypotheses.
\end{thebibliography}

\providecommand{\newblock}{}

\end{document}